\documentclass[12pt,letterpaper]{article}
\usepackage{jheppub}
\usepackage{verbatim}
\usepackage{slashed}

\usepackage{graphicx}
\usepackage{subfigure}
\input{epsf}
\usepackage{epsfig}
\usepackage{epstopdf}
\usepackage{amsthm}
\usepackage{tikz}

\def\({\left(} \def\){\right)}
\def\[{\left[} \def\]{\right]}

\newcommand{\be}{\begin{equation}}
\newcommand{\ee}{\end{equation}}
\newcommand{\bea}{\begin{eqnarray}}
\newcommand{\eea}{\end{eqnarray}}
\newcommand{\ba}{\begin{eqnarray}}
\newcommand{\ea}{\end{eqnarray}}

\newcommand{\beq}{\begin{equation}}
\newcommand{\eeq}{\end{equation}}
\newcommand{\beqa}{\begin{eqnarray}}
\newcommand{\eeqa}{\end{eqnarray}}
\newcommand{\beqar}{\begin{eqnarray*}}
\newcommand{\eeqar}{\end{eqnarray*}}

\newcommand{\eg}{{\it e.g.}\ }
\newcommand{\ie}{{\it i.e.}\ }

\title{Momentum in Single-trace $T\bar T$ Holography}

\author[a,b]{Soumangsu Chakraborty}
\author[c]{, Amit Giveon}
\author[d]{and  David Kutasov}

\affiliation[a]{Institut de Physique Th\'eorique, Universit\'e Paris-Saclay, CNRS, CEA\\	
Orme des Merisiers, 91191 Gif-sur-Yvette, France}
\affiliation[b]{Institute for Theoretical Physics, University of Amsterdam 	\\
1090GL Amsterdam, The Netherlands}
\affiliation[c]{Racah Institute of Physics, The Hebrew University, Jerusalem, 91904, Israel}
\affiliation[d]{Kadanoff Center for Theoretical Physics, Enrico Fermi Institute, and Department of Physics  \\University of Chicago, 5640 S. Ellis Ave, Chicago, IL 60637, USA}

\emailAdd{soumangsuchakraborty@gmail.com}
\emailAdd{giveon@mail.huji.ac.il}
\emailAdd{dkutasov@uchicago.edu}

\abstract{We extend the study of  \cite{Chakraborty:2020swe,Chakraborty:2023mzc} to black strings with general momentum, and discuss their interpretation in single-trace $T\bar T$ deformed $CFT_2$. 
}

\begin{document}
\maketitle
 
 \section{Introduction}\label{sec1}

Single-trace $T\bar T$ holography \cite{Giveon:2017nie} is obtained from  $AdS_3/CFT_2$ duality by adding to the Lagrangian of the $CFT_2$ a certain dimension $(2,2)$ quasi-primary operator $D(x^\alpha)$, $\alpha=0,1$, constructed in~\cite{Kutasov:1999xu}. This irrelevant deformation can be controlled, since in the dual bulk string theory on $AdS_3$ (supported by NS $B$-field), it corresponds to deforming the worldsheet theory, in which $AdS_3$ is described by the $SL(2,\mathbb{R})$ WZW model, by a truly marginal operator,~\cite{Giveon:2017nie}. This deformation has an Abelian Thirring form, and leads to a solvable theory. 

The resulting theory is often referred to as single-trace $T\bar T$ deformed CFT (see \eg~\cite{Giveon:2017nie,Araujo:2018rho,Chakraborty:2019mdf,Apolo:2019zai,Chakraborty:2020swe,Chakraborty:2020cgo,Chakraborty:2023mzc}), since it shares many properties with standard $T\bar T$ deformed CFT, \cite{Smirnov:2016lqw,Cavaglia:2016oda}: (1)~Both provide examples of irrelevant deformations of CFT's   that can be controlled; (2)~The operator $D(x^\alpha)$ has the same OPE's with the stress-tensor as $T\bar T$,~\cite{Kutasov:1999xu}; (3)~The spectrum of the single-trace $T\bar T$ deformed string theory is similar to that of $T\bar T$ deformed CFT. 

The similarity can be partially explained by the fact that the boundary theory corresponding to string theory on $AdS_3$ supported by NS $B$-field has much in common with symmetric product CFT.\footnote{In a class of constructions, \cite{Balthazar:2021xeh}, this correspondence is more precise.} In particular, the spectrum of long strings in that theory is the same as that of states with dimensions that go like $p^0$, in a CFT ${\cal M}^p/S_p$, with some seed CFT $\cal M$,~\cite{Argurio:2000tb,Giveon:2005mi}. The Abelian Thirring worldsheet deformation mentioned above acts on long string states as a $T\bar T$ deformation of $\cal M$,~\cite{Giveon:2017nie}. A similar comment appears to apply to BTZ black holes and their deformed version. 

The above observations led in recent years to a fruitful interplay between studies of single-trace $T\bar T$ deformed CFT and its (double-trace) cousin. In particular, it led to the solution of generalizations of $T\bar T$ deformed CFT,~\cite{Chakraborty:2018vja,Apolo:2018qpq,Chakraborty:2019mdf}, formulae for the partition sums of these theories,~\cite{Hashimoto:2019wct,Hashimoto:2019hqo}, and insights into their correlation functions,~\cite{Asrat:2017tzd,Giribet:2017imm,Cui:2023jrb}, and entanglement structure,~\cite{Chakraborty:2018kpr,Chakraborty:2020udr}. 

In  \cite{Chakraborty:2020swe,Chakraborty:2023mzc}, we studied situations where the deformed energies of states in single-trace $T\bar T$ deformed CFT become complex, from the bulk point of view. As in standard $T\bar T$ deformed CFT, this happens for low-energy states, like the $SL(2,\mathbb{R})$ invariant vacuum, for sufficiently large positive coupling, and for sufficiently high-energy states for negative coupling. We also discussed the relation between the size of the deformation of the boundary energy for states with energies that go like the central charge (like black-hole states and the $SL(2,\mathbb{R})$ invariant vacuum), and the corresponding deformed bulk geometry. 

In  \cite{Chakraborty:2020swe}, we showed that for negative coupling, the undeformed energy above which the deformed energy becomes complex is a maximal energy. In \cite{Chakraborty:2023mzc}, we showed that the (positive) coupling above which the deformed energy of the $SL(2,\mathbb{R})$ invariant vacuum becomes complex is a maximal coupling. 

In \cite{Chakraborty:2020swe,Chakraborty:2023mzc}, we focused on states with zero momentum on the boundary circle. The goal of this note is to generalize the discussion to non-zero momentum. In section~\ref{sec2},
we discuss black holes in the theory with positive single-trace $T\bar T$ coupling. In section~\ref{sec3}, we discuss the case of negative coupling.

 \section{Positive coupling}\label{sec2}

As in \cite{Chakraborty:2023mzc}, in this section, we will study type II string theory on $\mathbb{R}_t\times S^1_x\times T^4\times\mathbb{R}^4$, with $k$ NS fivebranes wrapping $S^1_x\times T^4$, and $p$ fundamental strings wrapping $S^1_x$. The ground state of the string-fivebrane system preserves eight supercharges, and has energy 
\begin{equation}\label{extE}
 E_{\rm ext}=\frac{kRv}{g^2l_s^2}+\frac{pR}{l_s^2}~,
 \end{equation}
where $R$ is the radius of the $x$ circle, and $v$ the volume of $T^4$ in string units, $V=(2\pi l_s)^4v$. 

We will focus on states living at the intersection of the strings and the fivebranes, $\mathbb{R}_t\times S^1_x$, that carry $n$ units of momentum\footnote{Without loss of generality, we will take $n>0$.} on the $x$ circle. The lowest energy states with non-zero $n$ preserve four supercharges and have energy 
\begin{equation}\label{ext}
 E_{\rm ext}^{(n)}=\frac{kRv}{g^2l_s^2}+\frac{pR}{l_s^2}+\frac{n}{R}~.
 \end{equation}

We will be interested in excited, non-extremal, states described by the black brane background, \cite{Maldacena:1996ky,Hyun:1997jv},
 \begin{equation}
 \begin{aligned}\label{ns5f1n}
 &ds^2=\frac{1}{f_1}\left[-\frac{f}{f_n}dt^2+f_n\left(dx-\frac{r_0^2\sinh2\alpha_n}{2f_n r^2}dt\right)^2\right]+f_5\left(\frac{1}{f}dr^2+d\Omega_3^2\right)+\sum_{i=1}^4 dx_i^4~,\\
 &e^{2\Phi}=g^2\frac{f_5}{f_1}~,\\
 &H=dx\wedge dt \wedge d\left(\frac{r_0^2\sinh2\alpha_1}{2f_1r^2}\right)+r_0^2\sinh2\alpha_5d\Omega_3~.
 \end{aligned}
 \end{equation}
Here $(x_1,..,x_4)$ are coordinates on $T^4$ and $(r,\Omega_3)$ are spherical coordinates on the transverse $\mathbb{R}^4$. The background \eqref{ns5f1n} has an event horizon at $r=r_0$ and an inner horizon at the origin, $r=0$. $g=e^{\Phi(r\to\infty)}$ is the string coupling far from the fivebranes; it is related to the ten-dimensional Newton constant via
 \begin{equation}\label{GNten}
 G_N^{(10)}=8\pi^6g^2l_s^8~.
 \end{equation}
 The harmonic functions in \eqref{ns5f1n}, $f$ and $f_{1,5,n}$, are given by 
 \begin{equation}\label{harfun}
 f=1-\frac{r_0^2}{r^2}~,\ \ \ \  f_{1,5,n}=1+\frac{r_{1,5,n}^2}{r^2}~, \ \ \ \ r_{1,5,n}^2=r_0^2\sinh^2\alpha_{1,5,n}~,
 \end{equation}
 with
 \begin{equation}\label{charges}
 \sinh2\alpha_1= \frac{2p g^2l_s^2}{vr_0^2}~, \ \ \ \   \sinh2\alpha_5=\frac{2l_s^2k }{r_0^2},\ \ \ \ \sinh 2\alpha_n=\frac{2l_s^4g^2n}{vR^2r_0^2}~.
 \end{equation}

The background \eqref{ns5f1n} is asymptotically flat; its ADM mass is \cite{Maldacena:1996ky}
 \begin{equation}\label{madm1}
 M_{\rm ADM}=\frac{Rvr_0^2}{2l_s^4g^2}(\cosh2\alpha_1+\cosh2\alpha_5+\cosh2\alpha_n)~.
 \end{equation}
The energy above the ground state \eqref{extE} is given  by
 \begin{equation}\label{eaext}
 E=M_{\rm ADM}-E_{\rm ext}=\frac{Rvr_0^2}{2l_s^4g^2}(e^{-2\alpha_1}+e^{-2\alpha_5}+\cosh2\alpha_n)~,
 \end{equation}
where we used \eqref{extE}, \eqref{charges}, \eqref{madm1}.

The entropy of the black brane \eqref{ns5f1n} is given by the Bekenstein-Hawking formula, $S=\frac{A}{4G_N}$,
\begin{equation}\label{ent1}
    S=\frac{L_xV_{S^3}V_{T^4}}{4G_{N}^{(10)}e^{2(\Phi-\Phi(\infty))}}\Bigg{|}_{r=r_0}=\frac{2\pi Rvr_0^3\sqrt{f_1f_5f_n} }{g^2l_s^4}\Bigg{|}_{r=r_0}.
\end{equation}
In the first equality, the numerator is the area of the event horizon. It is given by the product of the circumference of $S^1_x$, and the volumes of $S^3$ and $T^4$, evaluated with the metric \eqref{ns5f1n} at $r=r_0$: 
$L_x=2\pi R\sqrt{\frac{f_n}{f_1}}$, $V_{S^3}=2\pi^2(f_5r_0^2)^{3/2}$, $V_{T^4}=(2\pi)^4vl_s^4$. The denominator is $4G_N$ evaluated at the horizon. The second equality is obtained by plugging these expressions and the value of $G_{N}^{(10)}$, \eqref{GNten}, into the first one.

The decoupled theory on the fivebranes (known as Little String Theory (LST),~\cite{Aharony:1999ks,Kutasov:2001uf}) is obtained by taking the limit $g\to 0$, with $r$, $r_0$ and $r_{1,n}$ scaling like $g$, \cite{Aharony:1998ub}. To achieve that, we rescale 
 \begin{equation}\label{declim}
     r\to g r~, \ \ \ r_0\to g r_0~,  \ \ \ r_{1,n}\to g r_{1,n}
 \end{equation}
 in \eqref{ns5f1n}, and take 
 \begin{equation}\label{gto0}
     g\to 0
 \end{equation}
while holding the rescaled $r$, $r_0$ and $r_{1,n}$ fixed. 
Equations \eqref{harfun}, \eqref{charges} imply that in this limit $\alpha_5\to \infty$, and $f_5\to r_5^2/r^2$, with $r_5$ approaching the value
\begin{equation}\label{r5}
     r_5=\sqrt{k}l_s~.
 \end{equation}
The energy above the ground state \eqref{eaext} takes the form 
\begin{equation}\label{eextm3}
    E=\frac{Rvr_0^2}{2l_s^4}\left(e^{-2\alpha_1}+\cosh2\alpha_n\right),
\end{equation}
which generalizes eq. (2.12) in \cite{Chakraborty:2023mzc} to non-zero $n$, \eqref{charges}.

The black brane background \eqref{ns5f1n} is given in the limit \eqref{declim}, \eqref{gto0} by a product of a $T^4$, an $S^3$ of radius $r_5$, \eqref{r5}, with $k$ units of $H$ flux, described by a level $k$ supersymmetric $SU(2)$ WZW model, and a $2+1$ dimensional spacetime, with metric, dilaton and $H$-field 
\begin{equation}\label{ns5f1dec}
  \begin{aligned} 
  &ds^2=\frac{1}{f_1}\left[-\frac{f}{f_n}dt^2+f_n\left(dx-\frac{r_0^2\sinh2\alpha_n}{2f_n r^2}dt\right)^2\right]+\frac{r_5^2}{fr^2}dr^2~,\\
 &e^{2\Phi}=\frac{r_5^2}{f_1r^2}~,\\
 &H=dx\wedge dt \wedge d\left(\frac{r_0^2\sinh2\alpha_1}{2r_5^2}e^{2\Phi}\right)~,
 \end{aligned}
 \end{equation}
 where $f,f_{1,n}$ are given by \eqref{harfun}, \eqref{charges}, \eqref{declim}, 
 \begin{equation}\label{alpha1}
     \sinh 2\alpha_1=\frac{2l_s^2p}{vr_0^2},\ \ \ \ \sinh 2\alpha_n=\frac{2l_s^4n}{vR^2r_0^2}~.
 \end{equation}
As in \cite{Chakraborty:2023mzc}, it is convenient to change coordinates from $(t,r,x)$ in \eqref{ns5f1dec}, to $(\tau,\rho,\varphi)$, given by\footnote{We restrict here to the case $r_1>r_n$, or equivalently  $p>nl_s^2/R^2$. It is easy to generalize the discussion to the other regime.}
 \begin{equation}\label{coorbtz}
     \begin{aligned}
         \tau=\frac{r_5}{R}t ~,\ \ \ \ 
         \rho^2=\frac{R^2}{r_1^2-r_n^2}(r^2+r_n^2)~, \ \ \ \
         \varphi=\frac{x}{R}~.
     \end{aligned}
 \end{equation}
The rescaled spatial boundary coordinate has periodicity  $\varphi\sim\varphi+2\pi$, $r_5$ is given by \eqref{r5}, and $r_{1,n}$ are given in \eqref{harfun}, \eqref{alpha1}.
In the new coordinates, the background \eqref{ns5f1dec} takes the form\footnote{We chose the integration constant in going from $H$ to $B$ such that the $B$-field in \eqref{fullbgrho}, corresponding to the $H$-field in \eqref{ns5f1dec}, vanishes at $\rho=0$. This is compatible with the choice argued for in \cite{Ashok:2021ffx,Martinec:2023plo}.}
\begin{equation}\label{fullbgrho}
    \begin{aligned} 
        &ds^2=-\frac{N^2}{1+\frac{\rho^2}{R^2}}d\tau^2+\frac{d\rho^2}{N^2}+\frac{\rho^2}{1+\frac{\rho^2}{R^2}}\left(d\varphi-N_\varphi d\tau\right)^2~,\\
      &  B_{\tau\varphi}=\frac{\rho^2}{r_5}\sqrt{\left(1+\frac{\rho_-^2}{R^2}\right)\left(1+\frac{\rho_+^2}{R^2}\right)}~\frac{1}{1+\frac{\rho^2}{R^2}}~,\\
      & e^{2\Phi}= \frac{kv}{p}\sqrt{\left(1+\frac{\rho_-^2}{R^2}\right)\left(1+\frac{\rho_+^2}{R^2}\right)}~\frac{1}{1+\frac{\rho^2}{R^2}}~,
    \end{aligned}
\end{equation}
where
\begin{equation}\label{Nfull}
    N^2=\frac{(\rho^2-\rho_+^2)(\rho^2-\rho_-^2)}{r_5^2\rho^2}~, \ \ \ \ N_\varphi=\frac{\rho_+\rho_-}{r_5\rho^2}
\end{equation}
and, \eqref{coorbtz},
\begin{equation}\label{rho0full}
    \rho_+^2=\frac{r_0^2+r_n^2}{r_1^2-r_n^2}R^2~, \ \ \ \ \rho_-^2=\frac{r_n^2}{r_1^2-r_n^2}R^2~.
\end{equation}
It describes a rotating black string in $2+1$ dimensions, with an outer horizon at $\rho=\rho_+$ and an inner horizon at $\rho=\rho_-$, which correspond via \eqref{coorbtz} to $r=r_0$ and $r=0$, respectively. Upon reduction on the $\varphi$ circle, it gives rise to a two-dimensional black hole with  fundamental string winding and momentum $p$ and $n$. 

The background \eqref{fullbgrho}, \eqref{Nfull} 
approaches at large $\rho$ a linear dilaton spacetime; it describes a normalizable state in LST.\footnote{Or, more precisely, provides a thermodynamic description of a collection of such states, with the quantum numbers of the black brane.} For $R\gg\rho_\pm$, one can think of these states as living in the low energy theory. Indeed, taking $R\to\infty$ in \eqref{fullbgrho}, \eqref{Nfull}, we find the background
\begin{equation}\label{btxmetric}
 \begin{aligned}
   &  ds^2=-N^2d\tau^2+\frac{d\rho^2}{N^2}+\rho^2(d\varphi-N_{\varphi}d\tau)^2~,\\
 &  B_{\tau\varphi}=\frac{\rho^2}{r_5}~,\\
 &  e^{2\Phi}=\frac{kv}{p}~,  
 \end{aligned}
 \end{equation}
 with
 \begin{equation}\label{NNphi}
 \begin{aligned}
    & N^2=\frac{(\rho^2-\rho_+^2)(\rho^2-\rho_-^2)}{r_5^2\rho^2}\equiv\frac{\rho^2}{r_5^2}-8G_3M+\frac{(4G_3n)^2}{\rho^2}~,\\
    & N_{\varphi}=\frac{\rho_+\rho_-}{r_5\rho^2}\equiv\frac{4G_3 n}{\rho^2}~,
     \end{aligned}
 \end{equation}
which describes a BTZ black hole of mass $M$ and angular momentum 
 \begin{equation}\label{jn}
 J=n~,
 \end{equation}
 in an $AdS_3$ spacetime with $R_{\rm AdS}=r_5$, \eqref{r5}. The horizon positions $\rho_\pm$, and three-dimensional Newton constant, $G_3$, are given by
\begin{equation}\label{G3rho0}
   \rho_\pm^2=4 r_5G_3\left( r_5M\pm\sqrt{(r_5M)^2-J^2}\right)~, \ \ \ G_3=\frac{l_s}{4\sqrt{k}p}~.
\end{equation}
For finite $R$, the BTZ geometry is deformed at large $\rho$, where the background \eqref{fullbgrho}, \eqref{Nfull} approaches a linear dilaton one. The deformation is small near the horizon when $\rho_\pm\ll R$. Thus, the properties of such black holes are only slightly modified from the BTZ analysis. On the other hand, when this condition is not satisfied, the geometry \eqref{fullbgrho}, \eqref{Nfull} differs from its BTZ analog \eqref{btxmetric}, \eqref{NNphi} significantly for all $\rho$.

The above comments can be made quantitative by computing the energy $E$, \eqref{eextm3}, as a function of the charges $p$, $n$ and the size of the black hole. Using eqs. \eqref{harfun}, \eqref{eextm3}, \eqref{alpha1}, \eqref{rho0full}, one finds
\begin{equation}\label{mqlqr}
m\equiv E+\frac{Rp}{l_s^2}=\frac{2q_Lq_R}{q_L-q_R}\sqrt{\left(1+\frac{\rho_-^2}{R^2}\right)\left(1+\frac{\rho_+^2}{R^2}\right)}~,
\end{equation}
with 
\begin{equation}\label{qlr}
q_L=\frac{n}{R}-\frac{pR}{l_s^2}~,\ \ \ \ q_R=\frac{n}{R}+\frac{pR}{l_s^2}~.
\end{equation}
In eq. \eqref{mqlqr} we also defined a quantity $m$; one can show (see \eg\ \cite{Giveon:2005mi}) that it is equal to the ADM mass of the spacetime \eqref{fullbgrho} -- \eqref{rho0full}.

In deriving \eqref{mqlqr}, we have used the relations
\begin{equation}\label{qlqr}
    q_{L,R}=\frac{R^3r_0^2v}{l_s^4(\rho_+^2-\rho_-^2)}\left\{\frac{\rho_+\rho_-}{R^2} \mp \sqrt{\left(1+\frac{\rho_-^2}{R^2}\right)\left(1+\frac{\rho_+^2}{R^2}\right)}\right\}~,
\end{equation}
which follow from writing $n,p$ in \eqref{qlr} as functions of $\rho_\pm,r_0$, using \eqref{harfun}, \eqref{alpha1}, \eqref{rho0full},
\begin{equation}\label{np}
    \begin{aligned}
        &p=\frac{vR^2r_0^2}{l_s^2(\rho_+^2-\rho_-^2)} \sqrt{\left(1+\frac{\rho_-^2}{R^2}\right)\left(1+\frac{\rho_+^2}{R^2}\right)}~,\\
        &n=\frac{vR^2r_0^2}{l_s^4}~\frac{\rho_+\rho_-}{\rho_+^2-\rho_-^2}~.
    \end{aligned}
\end{equation}
The horizon radii, $\rho_\pm$, can be written in terms of the charges $m,q_{L,R}$ as
\begin{equation}\label{rhopm}
    \rho_{\pm}^2=-\frac{R^2}{2}\left[1+\frac{m^2}{q_Lq_R}\pm\frac{1}{q_Lq_R}\sqrt{(m^2-q_L^2)(m^2-q_R^2)}\right]~.
\end{equation}
The Bekenstein-Hawking entropy \eqref{ent1} of the black string \eqref{fullbgrho}, \eqref{Nfull} is
\begin{equation}\label{entropy}
  \begin{aligned}
    S&=\frac{2\pi R v\sqrt{k}}{l_s^3}\sqrt{(r_0^2+r_1^2)(r_0^2+r_n^2)}=\frac{2\pi kp}{r_5}\frac{\rho_+}{\sqrt{1+\frac{\rho_-^2}{R^2}}}\\
    &=\pi\sqrt{k}l_s\left(\sqrt{m^2-q_L^2}+\sqrt{m^2-q_R^2}\right)~,
    \end{aligned}
\end{equation}
where the first equality is obtained from \eqref{ent1} by taking the decoupling limit \eqref{declim} -- \eqref{r5}. The second equality in \eqref{entropy} is obtained by using \eqref{rho0full}, \eqref{np}, and the third equality is obtained by using \eqref{mqlqr}, \eqref{qlr}, \eqref{np}, \eqref{rhopm}. The last form of $S$ in \eqref{entropy} is well known; see \eg\ \cite{Giveon:2005mi,Giveon:2006pr}.

Comments: 
\begin{itemize}
    \item As mentioned above, $m$ is the ADM mass of the spacetime \eqref{fullbgrho} -- \eqref{rho0full}. It can be thought of as the ADM mass \eqref{madm1}, after subtracting the contribution due to the fivebranes to $E_{\rm ext}$ \eqref{extE}, which is infinite in the decoupling limit \eqref{gto0}.
    \item 
    When $\rho_\pm=0$, which is possible iff $n=0$, the background \eqref{fullbgrho} -- \eqref{rho0full} corresponds to the Ramond ground state in single-trace $T\bar T$ holography, \cite{Giveon:2017nie}. Its energy $E$ is zero for all $R$, as follows from \eqref{mqlqr}, \eqref{qlr}, in agreement with the fact that the state is half BPS.
    \item
    Quarter BPS states have $\rho_-=\rho_+>0$, which is possible iff $n\neq 0$. Their ADM mass is  $m=\frac{pR}{l_s^2}+\frac{n}{R}$, as follows from \eqref{qlr}, \eqref{rhopm}. It is equal to $E_{\rm ext}^{(n)}$ in \eqref{ext} after subtracting the (infinite) fivebrane contribution. The entropy \eqref{entropy} is $S=2\pi\sqrt{kpn}$ in these cases -- the well known three-charge supersymmetric black hole entropy~\cite{Maldacena:1996ky}.
    \item
    Geometries with $\rho_+>\rho_-$ in \eqref{fullbgrho}, \eqref{Nfull} correspond to excited states in single-trace $T\bar T$ holography. Their energy above the extremal energy is non-zero, and its value changes with $R$. 
    \item
    For large $R$, the entropy \eqref{entropy} takes the familiar $CFT_2$ form (with $c=6kp$),
    \begin{equation}\label{srinf}
    S= 2\pi \sqrt\frac{c}{12}\left(\sqrt{RE+n}+ \sqrt{RE-n}\right)~.
    \end{equation}
\end{itemize}
It is convenient to express the energy $E$, \eqref{mqlqr}, and entropy $S$, \eqref{entropy}, in terms of the quantities 
\begin{equation}\label{rhotilde}
    \tilde\rho_\pm\equiv\frac{\rho_\pm}{\sqrt{1+\frac{\rho_\mp^2}{R^2}}}~.
\end{equation}
Equation \eqref{np} implies that 
\begin{equation}\label{n/p}
n=\frac{p}{l_s^2}\tilde\rho_+\tilde\rho_-~.
\end{equation} 
The entropy \eqref{entropy} is given by
\begin{equation}\label{Sfixed}
S=\frac{2\pi kp}{r_5}{\tilde\rho_+}~,
 \end{equation}
and the energy \eqref{mqlqr} is
\begin{equation}\label{Erhotilde}
    E=\frac{Rp}{l_s^2}\left(-1+\sqrt{\left(1+\frac{\tilde{\rho}_-^2}{R^2}\right)\left(1+\frac{\tilde{\rho}_+^2}{R^2}\right)}\right).
\end{equation}
This equation can be obtained by substituting \eqref{qlr} in \eqref{mqlqr}, writing the resulting equation in terms of $\tilde{\rho}_{\pm}$ using \eqref{rhotilde}, and using \eqref{n/p} to eliminate~$n$. 

As in \cite{Chakraborty:2023mzc}, we can rewrite \eqref{Erhotilde} in terms of the single-trace $T\bar T$ coupling 
\begin{equation}\label{lambda}
    \lambda\equiv \frac{l_s^2}{R^2}~.
\end{equation}
For $\lambda=0$, the background \eqref{btxmetric} -- \eqref{G3rho0} describes a CFT state with energy and spatial momentum
\begin{equation}\label{E0}
     E(0)\equiv\frac{r_5}{R} M~,\ \ \ \ P\equiv\frac{n}{R}~.
 \end{equation}
As we vary $\lambda$, the energy of the state changes. To compute this variation, we need (as in \cite{Chakraborty:2023mzc}) to keep the entropy \eqref{Sfixed} fixed, and in the present discussion, we also need to keep the momentum $n$ \eqref{n/p} fixed. In other words, we need to impose on  \eqref{Erhotilde} the constraints
\begin{equation}\label{rholambda}
\tilde{\rho}_\pm(\lambda)=\tilde{\rho}_\pm(0)=\rho_\pm(0)~.
\end{equation} 
Imposing these constraints leads to 
\begin{equation}\label{zamo}
     \frac{1}{p}E(\lambda)=\frac{1}{\lambda R}\left(-1+\sqrt{1+2\lambda R\frac{E(0)}{p}+\left(\frac{\lambda RP}{p}\right)^2}\right)~,
 \end{equation}
which generalizes eq. (2.29) in \cite{Chakraborty:2023mzc} to non-zero momentum.

As in that paper, \eqref{zamo} is the same as what one would obtain by studying a symmetric product CFT ${\cal M}^p/S_p$. The generic state with large energy and momentum in the undeformed symmetric product CFT has (at large $E$, $P$, $p$) its energy and momentum equally split among the different copies of $\cal M$. Equation \eqref{zamo} gives the deformed energy of such states under a $T\bar T$ deformation of the seed CFT $\cal M$.

\section{Negative coupling}\label{sec3}

In \cite{Chakraborty:2020swe}, we studied the properties of single-trace $T\bar T$ deformed CFT for negative coupling. We restricted the discussion to vanishing spatial momentum ($n=0$ in \eqref{E0}). In this section, we will extend that discussion to general spatial momentum. 

The energy formula \eqref{zamo} has the property that for $\lambda<0$ and fixed $P$, there is a critical energy above which $E(\lambda)$ becomes complex. For $n=0$, this is reflected in the bulk geometry in the appearance of a naked singularity at some radial position \cite{Giveon:2017nie,Chakraborty:2020swe}. Our first goal is to find the geometry for $n>0$. This can be done 
\eg\ by boosting\footnote{See \eg \cite{Giveon:2006gw} for a description of the procedure.} 
the background in \cite{Chakraborty:2020swe}, and leads to:
\begin{equation}\label{lamneg}
  \begin{aligned} 
  &ds^2=\frac{1}{f_1}\left[-\frac{f}{f_n}dt^2+f_n\left(dx-\frac{r_0^2\sinh2\alpha_n}{2f_n r^2}dt\right)^2\right]+\frac{r_5^2}{fr^2}dr^2~,\\
 &e^{2\Phi}=\frac{r_5^2}{f_1r^2}~,\\
 &H=dx\wedge dt \wedge d\left(\frac{r_0^2\sinh2\alpha_1}{2r_5^2}e^{2\Phi}\right)~,
 \end{aligned}
 \end{equation}
 where $f,f_{1,n}$ are given by
 \begin{equation}\label{f1n}
     f=1-\frac{r_0^2}{r^2}~, \ \ \ f_1=-1+\frac{r_1^2}{r^2}~,\ \ \ f_n=1+\frac{r_n^2}{r^2}~, 
\end{equation}
with
\begin{equation}\label{ralpha}
    r_1^2=r_0^2\cosh^2\alpha_1~, \ \ \ r_n^2=r_0^2\sinh^2\alpha_n~,
\end{equation}
 and
 \begin{equation}\label{alphaneg}
     \sinh 2\alpha_1=\frac{2l_s^2p}{vr_0^2}~,\ \ \ \ \sinh 2\alpha_n=\frac{2l_s^4n}{vR^2r_0^2}~.
 \end{equation}
For $n=0$, this background reduces to that in \cite{Chakraborty:2020swe}.

Note the differences between the background \eqref{lamneg} -- \eqref{alphaneg} and the one in section \ref{sec2}:
 \begin{itemize}
\item 
The $1$ in the harmonic function $f_1$ in \eqref{harfun}, \eqref{ns5f1dec} is replaced by $-1$ in \eqref{lamneg},~\eqref{f1n}.
Consequently, as $r\to r_1$, we approach a curvature singularity that is not shielded by a horizon. The dilaton in \eqref{lamneg} diverges at the singularity as well.
\item
The relation between $r_1$ and $\alpha_1$ in \eqref{ralpha} differs from that in \eqref{harfun}. In particular, the location of the event horizon, $r=r_0$, is at a smaller radial distance from the inner horizon ($r=0$) than the location of the naked singularity, $r=r_1$, \ie\ $r_0\leq r_1$. The two coincide only for $\alpha_1\to 0$, \ie\ $r_0\to\infty$. 
\item
For $r<r_1$, the background \eqref{lamneg} -- \eqref{alphaneg} looks like a standard charged black-hole spacetime, with inner and outer horizons at $r=0$ and $r=r_0$, respectively. One can think of $r=r_1$ as a kind of UV cutoff.
\item
The background \eqref{lamneg} -- \eqref{alphaneg} can be obtained in the decoupling limit \eqref{declim} -- \eqref{r5} of the background in \eqref{ns5f1n} with the same harmonic functions $f,f_{5,n}$ as in \eqref{harfun}, \eqref{charges}, but with a different $f_1$, \eqref{f1n}, \eqref{ralpha}.
\end{itemize}
The Bekenstein-Hawking entropy of \eqref{lamneg} -- \eqref{alphaneg} associated with the horizon at $r=r_0$ is again given by \eqref{ent1}.  The ADM mass can be obtained by generalizing the analysis in \cite{Chakraborty:2020swe}, 
\begin{equation}\label{madmneg}
M_{\rm ADM}=\frac{Rvr_0^2}{2l_s^4g^2}(-\cosh2\alpha_1+\cosh2\alpha_5+\cosh2\alpha_n)~;
\end{equation}
the energy relative to the ground state is given by
\begin{equation}\label{eneg}
E=M_{\rm ADM}-E_{\rm ext}=\frac{Rvr_0^2}{2l_s^4g^2}(-e^{-2\alpha_1}+e^{-2\alpha_5}+\cosh2\alpha_n)~,
\end{equation}
where
\begin{equation}\label{extEneg}
E_{\rm ext}=\frac{kRv}{g^2l_s^2}-\frac{pR}{l_s^2}~,
\end{equation}
the energy of a BPS system of $k$ NS fivebranes and $p$ negative strings, \cite{Chakraborty:2020swe}.
After taking the fivebrane decoupling limit, one finds (compare to \eqref{eextm3})    
\begin{equation}\label{eneg1}
E=\frac{Rvr_0^2}{2l_s^4}(-e^{-2\alpha_1}+\cosh2\alpha_n)~.
\end{equation}
We can again change coordinates to
\begin{equation}\label{btzcoor}
    \tau=\frac{r_5}{R}t~,\ \ \ \rho^2=\frac{R^2}{r_1^2+r_n^2}(r^2+r_n^2)~, \ \ \ \varphi=\frac{x}{R}~,
\end{equation}
in terms of which the background \eqref{lamneg} -- \eqref{alphaneg} takes the form
\begin{equation}\label{rhogeometry}
    \begin{aligned}
        ds^2&= -\frac{N^2}{1-\frac{\rho^2}{R^2}}d\tau^2+ \frac{d\rho^2}{N^2}+\frac{\rho^2}{1-\frac{\rho^2}{R^2}}\left(d\varphi-N_{\varphi}d\tau\right)^2~,\\
        B_{\varphi\tau}&=\frac{\rho^2}{r_5}\sqrt{\left(1-\frac{\rho_-^2}{R^2}\right)\left(1-\frac{\rho_+^2}{R^2}\right)}\frac{1}{1-\frac{\rho^2}{R^2}}~, \\
        e^{2\Phi}&=\frac{kv}{p}\sqrt{\left(1-\frac{\rho_-^2}{R^2}\right)\left(1-\frac{\rho_+^2}{R^2}\right)}\frac{1}{1-\frac{\rho^2}{R^2}}~,
    \end{aligned}
\end{equation}
where
\begin{equation}\label{nnphineg}
    N^2=\frac{(\rho^2-\rho_+^2)(\rho^2-\rho_-^2)}{r_5^2\rho^2}~,  \ \ \ \ N_\varphi=\frac{\rho_+\rho_-}{r_5\rho^2}
\end{equation}
and, \eqref{ralpha},
\begin{equation}\label{rhopmneg}
    \rho_+^2=\frac{r_0^2+r_n^2}{r_1^2+r_n^2}R^2~, \ \ \ \rho_-^2=\frac{r_n^2}{r_1^2+r_n^2}R^2~.
\end{equation}
Note that equations \eqref{rhogeometry} -- \eqref{rhopmneg} can be formally obtained from \eqref{fullbgrho} -- \eqref{rho0full} by taking  $R^2\to -R^2$,  $r^2_1\to -r^2_1$. This is compatible with the fact that it describes the single-trace $T\bar T$ deformed CFT with negative coupling $\lambda$ \eqref{lambda}. 

In the limit $R\to \infty$, the coupling $\lambda\to 0$, and the background  \eqref{rhogeometry} -- \eqref{rhopmneg} approaches that of a BTZ black hole \eqref{btxmetric} -- \eqref{G3rho0}, a state in the undeformed CFT, as in section \ref{sec2}. For finite $R$, it has the following properties:
\begin{itemize}
    \item
    The IR and UV regimes, $\rho<R$ and $\rho>R$, are separated by a singularity  at $\rho=R$. This is the UV cutoff mentioned above. 
    \item
    Beyond the singularity, at $\rho>R$, the signature of $\varphi$ and $\tau$ in \eqref{rhogeometry} flips sign, as does $\exp(2\Phi)$. 
    \item
    For $\rho_+<R$, the asymptotic region 
    beyond the singularity, $\rho\gg R$, is a $2+1$ dimensional flat spacetime, consisting of a non-compact two-dimensional space with a linear dilaton, and a compact time, $\varphi\simeq\varphi+2\pi$. 
    \item
    For $\rho_+>R$,\footnote{The inner horizon is located at $\rho_-<R$, \eqref{rhopmneg}, regardless of the location of $\rho_+$.} the $B$-field and $\exp(2\Phi)$ in \eqref{rhogeometry} are imaginary for all $\rho$.
\end{itemize}

As discussed in \cite{Chakraborty:2020swe}, for $\rho_+<R$, the resulting background might lead to a sensible string theory. On the other hand, for $\rho_+>R$, the $B$-field is imaginary everywhere. Therefore, it is natural to interpret $\rho_+=R$ as a maximal size of the black hole, as suggested in \cite{Chakraborty:2023mzc} for the $SL(2,\mathbb{R})$ invariant vacuum. 
    
As mentioned above, in the $r$ coordinate of \eqref{lamneg} -- \eqref{alphaneg}, the location $r_1$ of the naked singularity increases when the location of the event horizon $r_0$ increases, and is always outside the black hole ($r_1>r_0$). The two approach each other as $r_0\to\infty$, as in \cite{Chakraborty:2020swe}. In this limit,
$r_n\to 0$, \eqref{ralpha}. In the $\rho$ coordinate, this limit corresponds to $\rho_+\to R$ and $\rho_-\to 0$, \eqref{rhopmneg}, the maximal value discussed above.

The ADM mass of the solution \eqref{rhogeometry} is given by
\begin{equation}\label{mdecneg}
    m=\frac{2q_Lq_R}{q_L-q_R}\sqrt{\left(1-\frac{\rho_-^2}{R^2}\right)\left(1-\frac{\rho_+^2}{R^2}\right)}~,
\end{equation}
the analog of \eqref{mqlqr} for negative $\lambda$ \eqref{lambda}.
The charges $q_{L,R}$ are
\begin{equation}\label{qlrneg}
    q_L=\frac{n}{R}+\frac{pR}{l_s^2}, \ \ \ \ q_R=\frac{n}{R}-\frac{pR}{l_s^2}~,
\end{equation}
which is obtained from \eqref{qlr} by taking $p\to -p$, or exchanging $q_L$ and $q_R$. 

To derive \eqref{mdecneg}, we recall that
\begin{equation}\label{EEneg}
m=E-\frac{Rp}{l_s^2}~,
\end{equation}
the negative string analog of \eqref{mqlqr}, 
$E$ is given by eq. \eqref{eneg1}, 
and we have used the relations
\begin{equation}\label{qlqrneg}
    q_{L,R}=\frac{R^3r_0^2v}{l_s^4(\rho_+^2-\rho_-^2)}\left\{\frac{\rho_+\rho_-}{R^2} \pm \sqrt{\left(1-\frac{\rho_-^2}{R^2}\right)\left(1-\frac{\rho_+^2}{R^2}\right)}\right\}~,
\end{equation}
which follow from writing $n,p$ as a function of $\rho_\pm,r_0$ (using \eqref{ralpha}, \eqref{alphaneg}, \eqref{rhopmneg}),
\begin{equation}\label{npneg}
    \begin{aligned}
        &p=\frac{vR^2r_0^2}{l_s^2(\rho_+^2-\rho_-^2)} \sqrt{\left(1-\frac{\rho_-^2}{R^2}\right)\left(1-\frac{\rho_+^2}{R^2}\right)}~,\\
        &n=\frac{vR^2r_0^2}{l_s^4}~\frac{\rho_+\rho_-}{\rho_+^2-\rho_-^2}~.
    \end{aligned}
\end{equation}
 The horizon radii, when expressed in terms of $m,q_{L,R}$, take the form
\begin{equation}\label{negrhopm}
    \rho_{\pm}^2=\frac{R^2}{2}\left[1+\frac{m^2}{q_Lq_R}\mp\frac{1}{q_Lq_R}\sqrt{(m^2-q_L^2)(m^2-q_R^2)}\right]~.
\end{equation} 

The Bekenstein-Hawking entropy \eqref{ent1} of the solution \eqref{rhogeometry} -- \eqref{rhopmneg} is
\begin{equation}\label{entropyneg}
  \begin{aligned}
    S&=\frac{2\pi R v\sqrt{k}}{l_s^3}\sqrt{(-r_0^2+r_1^2)(r_0^2+r_n^2)}=\frac{2\pi kp}{r_5}\frac{\rho_+}{\sqrt{1-\frac{\rho_-^2}{R^2}}}\\
    &=\pi\sqrt{k}l_s\left(\sqrt{q_L^2-m^2}+\sqrt{q_R^2-m^2}\right)~,
    \end{aligned}
\end{equation}
where the first equality is obtained from \eqref{ent1} with $f_1$ given by \eqref{f1n}, \eqref{ralpha} in  the decoupling limit described around equations \eqref{declim} -- \eqref{r5}. The second equality in \eqref{entropyneg} is obtained by using \eqref{rhopmneg}, \eqref{npneg}, and the third equality is obtained by using \eqref{mdecneg}, \eqref{qlrneg}, \eqref{npneg}, \eqref{negrhopm}. 
The entropy \eqref{entropyneg} takes its maximal value when the event horizon approaches its maximal value, $\rho_+\to R$, with $\rho_-\to 0$ in the limit.\footnote{Note that $\rho_-\to 0$ when $\rho_+\to R$ for fixed $n$; this follows from \eqref{ralpha}, \eqref{alphaneg}, \eqref{rhopmneg}.}

The inverse temperature, $\beta$, is given by
\begin{equation}\label{beta}
    \beta=\frac{2\pi r_5}{r_0^2}\sqrt{(-r_0^2+r_1^2)(r_0^2+r_n^2)}=2\pi R\sqrt{1-\frac{\rho_+^2}{R^2}}\frac{r_5\rho_+}{\rho_+^2-\rho_-^2}~.
\end{equation}
It diverges when the event horizon approaches its maximal value, $\rho_+\to R$. 

As in section \ref{sec2}, we can compute the energy \eqref{mdecneg} -- \eqref{EEneg} in terms of $\rho_\pm$. Repeating the same steps as there, we have the same expressions for $n$ and $S$ as in \eqref{n/p} and \eqref{Sfixed}, respectively, but with $\tilde{\rho}_{\pm}$ given by 
\begin{equation}\label{rhotildeneg}
    \tilde{\rho}_{\pm}\equiv\frac{\rho_\pm}{\sqrt{1-\frac{\rho_\mp^2}{R^2}}}~,
\end{equation}
rather than \eqref{rhotilde}. The energy \eqref{EEneg} is now given by 
\begin{equation}\label{Erhotildeneg}
    E=-\frac{Rp}{l_s^2}\left[-1+\sqrt{\left(1-\frac{\tilde{\rho}_-^2}{R^2}\right)\left(1-\frac{\tilde{\rho}_+^2}{R^2}\right)}~\right].
\end{equation}
Defining the coupling 
\begin{equation}\label{lambdaneg}
    \lambda\equiv- \frac{l_s^2}{R^2}<0~,
\end{equation}
and demanding that the momentum $n$ and entropy $S$ do not change with $\lambda$, leads again to the constraint \eqref{rholambda}. Plugging this into the expression for the energy, we find \eqref{zamo} again, this time for negative coupling \eqref{lambdaneg}. 

For $\rho_\pm=0$, the background \eqref{rhogeometry}, \eqref{nnphineg} corresponds to the Ramond ground state of the theory.  It is a $1/2$ BPS state, with $E=0$ for all $R$, \eqref{Erhotildeneg}.  
$\rho_+=\rho_->0$ corresponds to $1/4$ BPS states with $m=q_R$, \ie\ $E=n/R$, $S=2\pi\sqrt{kpn}$, \eqref{entropyneg}, and $T=0$, \eqref{beta}.
    
Equation \eqref{zamo} has the property that for a given $\lambda<0$ and $P$ (or $n$, \eqref{E0}), there is a maximal undeformed energy $E(0)$ beyond which the energy $E(\lambda)$ becomes complex. If we demand that we stay below this maximal energy, the deformed energy remains below a critical energy, 
\begin{equation}\label{Ec}
E(\lambda)< E_c=\frac{pR}{l_s^2}~.   
\end{equation}
The region of $E(0)$ where $E(\lambda)$ is complex corresponds precisely to geometries \eqref{rhogeometry} -- \eqref{rhopmneg} in which the $B$-field and dilaton are complex. This can be seen by noting that  
\begin{equation}\label{sqrt}
\sqrt{1+2\lambda R\frac{E(0)}{p}+\left(\frac{\lambda RP}{p}\right)^2}=\left[1-\left(\frac{\lambda RP}{p}\right)^2\right]\sqrt{\left(1-\frac{\rho_-^2}{R^2}\right)\left(1-\frac{\rho_+^2}{R^2}\right)}~,
\end{equation}
which follows from \eqref{mdecneg} -- \eqref{EEneg},  \eqref{Erhotildeneg}. The l.h.s. of \eqref{sqrt} appears in the formula for $E(\lambda)$, \eqref{zamo}, while the second factor on the r.h.s. appears in the expression \eqref{rhogeometry} for the supergravity fields. Thus, they are either both real or both imaginary.  

We conclude that the analysis of \cite{Chakraborty:2020swe}, that was done for $n=0$, generalizes to non-zero $n$. In particular, one can check that there is a detailed match between the geometry \eqref{fullbgrho} -- \eqref{rho0full}, and \eqref{rhogeometry} -- \eqref{rhopmneg}, and the corresponding energies \eqref{zamo} and entropies \eqref{entropy}, \eqref{entropyneg}. For negative $\lambda$, the critical energy $E_c$  \eqref{Ec} should be viewed as a maximal energy for all $n$. It is natural to conjecture that an analogous statement is true for standard $T\bar T$ deformed CFT. 

It is interesting to ask what is the geometry that corresponds to the maximal energy \eqref{Ec}. It follows from \eqref{sqrt} that it corresponds to setting $\rho_+=R$ and, due to \eqref{n/p}, \eqref{rhotildeneg}, $\rho_-=0$. In this case, the $B$-field in \eqref{rhogeometry}, and $N_\varphi$ in \eqref{nnphineg} vanish, and the dilaton $\Phi$ requires an infinite shift to remain finite. The resulting geometry is universal (in the sense that it does not depend on the particular value of $n$). We will defer a more detailed study of it to future work.

\section*{Acknowledgements} 

The work of SC received funding under the Framework Program for Research and
“Horizon 2020” innovation under the Marie Skłodowska-Curie grant agreement n° 945298. 
The work of AG and DK is supported in part by the BSF (grant number 2018068).
The work of AG is supported in part by the ISF (grant number 256/22). The work of DK is supported in part by DOE grant DE-SC0009924. This work was supported in part by the FACCTS Program at the University of Chicago.

 \newpage


\providecommand{\href}[2]{#2}\begingroup\raggedright\endgroup

\end{document}